\documentclass[aps,pra,twocolumn,amsfonts,amssymb,amsmath,showpacs,
floatfix,nofootinbib,groupedaddress,superscriptaddress,citesort]{revtex4}
\usepackage{mathrsfs}
\usepackage{amsfonts}
\usepackage{amstext}
\usepackage{amsmath}
\usepackage{amssymb}
\usepackage{bm}
\usepackage{bbm}
\usepackage[dvips]{graphicx}
\def\qed{\leavevmode\unskip\penalty9999 \hbox{}\nobreak\hfill
     \quad\hbox{\leavevmode  \hbox to.77778em{%
              \hfil\vrule   \vbox to.675em%
               {\hrule width.6em\vfil\hrule}\vrule\hfil}}
     \par\vskip3pt}

\begin{document}

\title{Incoherent Gaussian equivalence of $m-$mode Gaussian states}

\author{Shuanping Du}
\email{dushuanping@xmu.edu.cn} \affiliation{School of Mathematical
Sciences, Xiamen University, Xiamen, Fujian, 361000, China}

\author{Zhaofang Bai}\thanks{Corresponding author}
\email{baizhaofang@xmu.edu.cn} \affiliation{School of Mathematical
Sciences, Xiamen University, Xiamen, Fujian, 361000, China}

\begin{abstract}

Necessary and sufficient conditions for arbitrary multimode (pure or mixed) Gaussian states to be equivalent under incoherent Gaussian operations are derived. We show that two Gaussian states are incoherent  equivalence if and only if they are related by incoherent unitaries. This builds the counterpart of the celebrated result that two pure entangled states are equivalent under LOCC if and only if they are related by local unitaries. Furthermore, incoherent equivalence of Gaussian states is equivalent to frozen coherence [Phys. Rev. Lett. \textbf{114}, 210401 (2015)].
Basing this as foundation, we find all measures of coherence are frozen for an initial Gaussian state under strongly incoherent Gaussian operations if and only if the relative entropy measure of coherence is frozen for the state. This gives an entropy-based dynamical condition in which the coherence of an open quantum system is totally unaffected by noise.

\end{abstract}

\maketitle

{\it \bf I. Introduction}

\vspace{0.1in}
Quantum coherence, being at the heart of interference phenomena, stands as one of intrinsic features of quantum mechanics that induces a number of intriguing phenomena in quantum optics \cite{Glau1, Scul, Albre1, Walls1} and quantum information \cite{Nile1}. It  constitutes a powerful resource for quantum computing \cite{Hill}, cryptography \cite{Cole}, information processing \cite{Stre1,An,Diaz}, thermodynamics \cite{Lost}, metrology \cite{Frow}, and quantum biology \cite{Stre3}.

The first framework for understanding quantum coherence is quantum optics which require quantum states in a continuous-variable system.
Gaussian states that are processed in most optical experiments have arisen to a privileged position in the context of quantum computation \cite{Weed, Terhal, Grim} over different physical platforms, such as optical \cite{Pfister}, trapped ions \cite{Serafini,Fluhmann}, atomic ensembles \cite{Milne, Motes} and hybrid systems \cite{Aolita}.
 Recently, there is a growing interest in building resource theory of coherence of Gaussian states \cite{Xu1, Hou1, Giulio, Du1, Hahn}.
Coherence effects of Gaussian states have
been addressed in different branches of quantum information, for example, unitary process in quantum thermodynamics \cite{Camati}, coherence as a resource in charging quantum battery \cite{GPHC}, dynamical behavior of quantum coherence of a displaced squeezed thermal state \cite{Ali}. In view of resource theory \cite{Ben1,Vidal1,Wolf},
one fundamental question is to classify coherent Gaussian states.  A natural way of defining equivalence relations in the set of coherent Gaussian states is that equivalent states contain the
same amount of coherence. Since the primary tool for
analyzing coherent Gaussian states is incoherent Gaussian operations which are powerful to describe noise and decoherence
of optical systems \cite{Xu1,Xu2,San,WGD},
the monotonicity of coherence under incoherent Gaussian operations leads to that we can identifying any two states
which can be transformed from each other with certainty by incoherent Gaussian operations. Clearly, this criterion is interesting in quantum information theory since equivalent states are indistinguishable for exactly the same tasks. The aim of this note is to characterize equivalence of coherent Gaussian states under incoherent Gaussian operations.

For entangled states, a beautiful result is that two pure states can be transformed from each other  with certainty by LOCC if and only if they are related by local unitaries \cite{Ben1, Vidal1}. The equivalence of entangled Gaussian states under local unitaries is discussed for bipartite setting \cite{Wolf} and for more parties \cite{Adesso,Wang,Ade2,Ser,Gie}. In \cite{Ade2,Ser}, standard forms of generic $m-$mode pure and mixed states are introduced. For generic pure Gaussian states, it is shown that Gaussian local unitaries equivalence classes are classified by three positive numbers related to local purities. The case of $3-$mode has been discussed detailly in \cite{Ade3}. We will classify coherent Gaussian states in terms of incoherent unitaries and  the relative entropy measure of coherence.

The paper is organized as follows. In section II,  we present an explicit description of coherent Gaussian states and incoherent Gaussian operation.
In section III, we present the results for incoherent Gaussian equivalence of $m-$mode Gaussian states that we addressed. Section IV is a summary of our findings. Appendix is the proof of our results.

\vspace{0.1in}

{\it \bf II. Background and notation}
\vspace{0.1in}

Here, we provide some background on bosonic Gaussian states and Gaussian operations (see \cite{Weed} for a review).
Let ${\mathcal H}$ be
an infinite dimensional Hilbert space with fixed Fock basis $\{|n\rangle\}_{n=0}^{+\infty}$.
When we consider the $m$-mode continuous-variable systems ${\mathcal H}^{\otimes m}$, we adopt $(\{|n\rangle\}_{n=0}^{+\infty})^{\otimes m}$
as its reference basis.
For a quantum state $\rho\in {\mathcal H}^{\otimes m}$, the characteristic function of $\rho$ is defined as
$$\begin{array}{lll}
{\mathcal X}_\rho(\lambda) & = & tr(\rho D(\lambda))\\
D(\lambda) & = & \otimes _{i=1}^m D(\lambda_i)\\
D(\lambda_i) & = & e^{(\lambda_i\widehat{a_i}^\dag-{\overline \lambda_i}\widehat{a_i})},
\end{array}$$
here $\widehat{a_i}$ and $\widehat{a_i}^\dag$ are the annihilation operator and the creation operator in mode $i$, \quad $\lambda=(\lambda_1, \cdots, \lambda_m)^t, \quad  {\overline \lambda_i}$ denotes the complex conjugate of $\lambda_i$. Gaussian states are those states for which ${\mathcal X}_\rho(\lambda)$  is a Gaussian function of the phase space, i.e.,
$${\mathcal X}_\rho(\lambda)=exp^{-\frac{1}{4}{\overrightarrow r}\Omega V\Omega^t{\overrightarrow r}^t-i(\Omega d)^t{\overrightarrow r} ^t},$$ where $\overrightarrow{r}=(\lambda_{1x}, \lambda_{1y}, \ldots,\lambda_{mx}, \lambda_{my} )$, $\lambda_{jx},\lambda_{jy}$ are the real part and imaginary part of $\lambda_j \ (j=1,2, \ldots, m)$, V is a $2m\times 2m$ real hermitian matrix which is called covariance matrix satisfying the uncertainty relation $V+i\Omega\geq 0$, $d\in {\mathbb R}^{2m}$ is called mean value,
$\Omega=\bigoplus_{k=1}^m \left(\begin{array}{cc}
                                  0 & 1\\
                                  -1 & 0\end{array}\right)$ \cite{Weed}. Note that $\det V \geq 1$ and $\det V = 1$ if and only if $\rho$ is pure. It is clear that $(V,d)$
can describe Gaussian state $\rho$ completely. So $\rho$ can be usually written in $\rho(V, d)$.
Every Gaussian operation is a completely positive trace-preserving mapping that takes Gaussian states
to Gaussian states. It
is described by $(T, N, {\overline d})$, it performs on $\rho(V,d)$ and gets
Gaussian state with mean value and covariance matrix as follows:  $$ d\mapsto Td+{\overline d}, \hspace{0.1in} V\mapsto TVT^t+N,$$ where
${\overline d}\in{\mathbb R}^{2m}$, $T, N$ are $2m\times 2m$ real matrices with $N+i\Omega-iT\Omega T^t\geq 0$ (complete positivity condition)  \cite{Weed,Holex}.

Inspired by the idea of discrete-variable systems \cite{Bau}, the framework for quantifying coherence of Gaussian states has been built in \cite{Xu1,Xu2}.
The incoherent Gaussian states are defined as diagonal Gaussian states in Fock basis. Every incoherent  Gaussian state has the form $\otimes_{i=1}^m\rho_{th}^{A_i}(\overline{n_i})$, here
$\rho_{th}^{A_i}({\overline n_i})=\sum_{n=0}^{+\infty} \frac{{\overline n_i}^n}{({\overline n_i}+1)^{n+1}}|n\rangle\langle n|$ is the incoherent states of $i$th-mode $A_i$. The set of incoherent Gaussian states will be
labelled by ${\mathcal I}$. A Gaussian operation is incoherent if it maps  incoherent Gaussian states into incoherent Gaussian states. In fact,
a Gaussian operation $\Phi(T, N, {\overline d})$ is incoherent if and only if $$\left.\begin{array}{lll}
                                                                     {\overline d} & = & 0,\\
                                                                      T & =& \{t_jO_j\}_{j=1}^m\in {\mathcal T}_{2m},\\
                                                                      N & = & \oplus_{j=1}^m \omega_j I_2,\\
                                                                      \omega_j &\geq &|1-\sum_{k, r(k)=j} t_k^2\det O_k|,\forall j,\end{array}\right.$$
where $t_j, \omega_j\in {\mathbb R}$, $O_j$ is a $2\times 2$ real orthogonal matrix $(O_jO_j^t=I_2)$, ${\mathcal T}_{2m}$ denotes the set of $2m\times 2m$ real matrices such that, for any $T\in {\mathcal T}_{2m}$, the $(2j-1,2j)$ two columns of $T$ have just one $2\times 2$ real matrix $t_jO_j$ located in $(2r(j)-1, 2r(j))$ rows for $\forall j$, $r(j)\in
\{k\}_{k=1}^m$, and other elements are all zero. For Gaussian state $\rho(V,d)$, it performs on $\rho(V,d)$ and obtains
Gaussian state with mean value and covariance matrix as follows:  $$ d\mapsto Td, \hspace{0.1in} V\mapsto TVT^t+N.$$
Specially, unitary operators of ${\mathcal T}_{2m}$ are called  incoherent unitaries in this note.

Based on the definition of incoherent Gaussian states and incoherent Gaussian operations ($\text{IGO}s$),
any proper coherence measure $C$ is a non-negative function and
must satisfy the following
conditions \cite{Xu1}:

$({\text C1})$ $C(\rho)=0$ for all $\rho\in{\mathcal I}$;

$({\text C2})$ Monotonicity under all incoherent Gaussian operations ($\text{IGO}s$) $\Phi$:

$$C(\Phi(\rho))\leq C(\rho);$$

$({\text C3)}$ Non-increasing under mixing of Gaussian states:
$$C(\sum_jp_j\rho_j)\leq \sum_jp_jC(\rho_j),$$ for any set of Gaussian states
$\{\rho_j\}$ and any $p_j\geq 0$ with $\sum_jp_j=1$. Note that the set of Gaussian states is not convex, thus $\rho_j$ and $\sum_jp_j\rho_j$ in
$({\text C3})$ should be all Gaussian.
Based on the definition of the coherence measure, the relative entropy measure has been provided by $$C_R(\rho)=\inf _{\delta\in {\mathcal I}} S(\rho||\delta),$$ $S(\rho||\delta)=tr(\rho\log_2\rho)-tr(\rho\log_2\delta)$ being the relative entropy. Furthermore
$$C_R(\rho)=-S(\rho)+\sum\limits_{i=1}^m[({\overline n_i}+1)\log_2({\overline n_i}+1)-n_i\log_2 {\overline n_i}],$$
$S(\rho)=\sum\limits_{i=1}^m[\frac{v_i-1}{2}\log_2\frac{v_i-1}{2}-\frac{v_i+1}{2}\log_2 \frac{v_i+1}{2}]$,
${\overline n_i}=\frac{1}{4}[tr(V^{(i)})+\|d^{(i)}\|^2-2]$, where $S(\rho)$ is the von Neumann entropy of $\rho$,
$\{v_i\}_{i=1}^m$ are symplectic eigenvalues of $V$ \cite{Ade6}, and ${\overline n_i}$ is determined by $i$th-mode
covariance matrix $V^{(i)}$ and mean value $d^{(i)}$, $\|d^{(i)}\|$ is the Euclidean norm of $d^{(i)}$ \cite{Xu1}.

For coherent Gaussian states $\rho(V,d),\quad \sigma(V', d')$, we say they are IGO equivalence if there exist IGOs $\Phi$ and $\Psi$ satisfying
$\Phi(\rho(V, d))=\sigma(V', d')$ and $\Psi(\sigma(V',d'))=\rho(V, d)$. We denote it by $\rho(V,d)\stackrel{\text IGO}{\sim}\sigma(V',d')$. We will provide a structural characterization of equivalence for coherent Gaussian states by incoherent unitary operations and the coherence measure based on the relative entropy.
\vspace{0.1in}

{\it \bf III. Main results}
\vspace{0.1in}

Before giving our main results, we need to introduce the concept of strictly incoherent Gaussian operations which is originated from the definition of strictly incoherent operations of discrete-variable systems \cite{Win2, Gour2}. It plays a key role in classifying IGO equivalence of Gaussian states.

{\bf Definition 3.1.} An incoherent Gaussian operation $\Phi(T,N)$ is called strictly incoherent if
 each $(2i-1, 2i)$ row and each $(2j-1, 2j)$ column of $T$ has just one element of $\{t_jO_j\}^m_{ j=1}$, $\omega_j \geq |1-t_k^2\det O_k|\ (r(k)=j)$.

Throughout the note, we write mean values and covariance matrices of $m-$mode Gaussian states in terms of two dimensional subblocks as
$$d=(d_1, d_2, \cdots, d_m)^t,V=\left(\begin{array}{cccc}
                             V_{11}& V_{12}& \cdots & V_{1m}\\
                             V_{12}^t & \ddots & \ddots & \vdots\\
                             \vdots & \ddots & \ddots & V_{m-1, m}\\
                             V_{1m}^t & \cdots & V_{m-1, m}^t & V_{mm}\end{array}\right).$$  $V_{ii}$ and $d_i$ are the $i$th-mode
covariance matrix and mean value respectively.
It is known that each subvector $d_i$ and off-diagonal block $V_{ij}$  become $0$ for every incoherent Gaussian state \cite{Xu1}. The nonzero elements of off-diagonal blocks of $V$ and subvectors of $d$ reveal coherence of Gaussian states. In order to classify coherent Gaussian states, we assume that every row of $V$ has at least one nonzero off-diagonal block.

Now, we are in a position to give our main results.

{\bf Theorem 3.2.} {\it Assume  $m\geq 2$ and every row of $V$ and $V'$ have at least one nonzero off-diagonal block, then the
following statements are equivalent:

(i) $\rho(V,d)\stackrel{\text IGO}{\sim}\sigma(V',d')$;

(ii) There exists an incoherent Gaussian unitary operator $U$ such that $UVU^t=V', Ud=d'$, here  $U=P_{\pi}\otimes I_2 \left(\begin{array}{ccc}
                           O_1 & & 0\\
                               &\ddots &\\
                               0& & O_m\end{array}\right)$,  $P_{\pi}$ is the permutation matrix corresponding to a permutation $\pi$ of $\{1,2,\cdots, m\}$,
$O_i$ $(1\leq i\leq m)$ are $2\times 2$ real orthogonal matrices with $\det O_i=1$, $I_2$ is the $2\times 2$ unit matrix;

(iii) There is a strictly incoherent Gaussian operation $\Phi$ satisfying $\Phi(\rho(V,d))=\sigma(V',d')$ and $C_R(\rho(V,d))=C_R(\sigma(V',d'))$.}

For one-mode case, our result reads as follows.

{\bf Theorem 3.3.} {\it Assume  $d\neq 0$ or $V\neq \lambda I_2$\ ($\rho(V,d)$ is coherent),  $I_2$ is the $2\times 2$ unit matrix, then the
following statements are equivalent:

(i) $\rho(V,d)\stackrel{\text IGO}{\sim}\sigma(V',d')$;

(ii) There exists an incoherent Gaussian unitary operator $U$ such that $UVU^t=V', Ud=d'$, here  $U= \left(\begin{array}{cc}
                           \cos\theta  & \sin\theta\\
                               -\sin\theta & \cos\theta\end{array}\right)$ for some $\theta\in{\mathbb R}$;

(iii) There is an incoherent Gaussian operation $\Phi$ satisfying $\Phi(\rho(V,d))=\sigma(V',d')$ and $C_R(\rho(V,d))=C_R(\sigma(V',d'))$.}

\vspace{0.1in}


 Theorem 3.2 and 3.3 show the existence of measure-independent
freezing of coherence. Quantum coherence is a useful physical resource, but coherence of a Gaussian state is decreasing under IGOs $(C(\Phi(\rho))\leq C(\rho))$.
The loss of coherence may weaken the abilities of a Gaussian state to fulfil certain quantum information processing tasks.
An interesting question is to study when the coherence of an open system is frozen \cite{Ade4}, i.e., when $C(\Phi(\rho))=C(\rho)$ for a coherent Gaussian state $\rho$ and an IGO $\Phi$.
However, some coherence measures being frozen do not imply other coherence measures being frozen too, since
different coherence measure results in different orderings of coherence in general \cite{Tong1}. Freezing of coherence is dependent on the coherence measures in general.
By Theorem 3.2 and 3.3, we find all measures of coherence are frozen for an input Gaussian state
 if and only if the relative entropy measure of coherence is frozen for the state. A parallel result in discrete-variable systems is that
all measures of coherence are frozen for an initial state in a strictly incoherent channel if and only if
the relative entropy of coherence is frozen for the state \cite{Ma2}.

Theorem 3.2 and Theorem 3.3 present frozen  phenomenon of coherence and entanglement simultaneously \cite{Raji}.
An important class of 2-mode Gaussian states has covariance matrices in standard form
$$V=\left(\begin{array}{cc}
                aI_2 & C\\
                C & bI_2\end{array}\right), \quad
C=\left(\begin{array}{cc}
                c & 0\\
                0 & d \end{array}\right),$$ $a\geq 1, b\geq 1, c,d\in{\mathbb R}$ \cite{Weed, Duan1, Simon1}. Any Gaussian state can be transformed to the
Gaussian state with the covariance matrix in standard form by local linear unitary Bogoliubov operations \cite{Duan1}. Let $d=(d_1, d_2)$, $\rho(V,d)^\sharp$ denote the set of all Gaussian states which are incoherent equivalence with $\rho(V,d)$, by Theorem 3.2,
$$\rho(V,d)^\sharp=\left\{\sigma(V_1,(O_1d_1, O_2d_2)^t)\right\}\cup \left\{\delta(V_2,(O_1d_2, O_2d_1)^t) \right\},$$
$V_1=\left(\begin{array}{cc}
                aI_2 & O_1CO_2^\dag\\
                O_2CO_1^\dag & bI_2\end{array}\right),$ $V_2=\left(\begin{array}{cc}
                bI_2 & O_1CO_2^\dag\\
                O_2CO_1^\dag & aI_2\end{array}\right),$
$O_i=\left(\begin{array}{cc}
                \cos\theta_i & \sin\theta_i\\
                -\sin\theta_i & \cos\theta_i\end{array}\right)$, $\forall \ \theta_i\in{\mathbb R}, i=1,2$.
 Note that any transformation of $$V\mapsto  \left(\begin{array}{cc}
                O_1 & 0\\
                0  & O_2\end{array}\right)V\left(\begin{array}{cc}
                O_1^t & 0\\
                0  & O_2^t\end{array}\right)=V_1$$ is a special kind of  local linear unitary Bogoliubov operations \cite{Duan1}, this tells that
 the amount of coherence of $\left\{\sigma(V_1,(O_1d_1, O_2d_2)^t)\right\}$
 and the amount of entanglement of Gaussian states of $\left\{\sigma(V_1,(O_1d_1, O_2d_2)^t)\right\}$ are equal, respectively. In addition,
 by a direct computation, the symplectic spectrum $\{v_+,\ v_-\}$ of $V_i$ $(i=1,2)$ are given by  $$v_{\pm}=\sqrt{\frac{\Delta\pm\sqrt{\Delta^2-4\det V}}{2}},$$ $\Delta=a^2+b^2+2\det C$ (one can also refer to \cite{Weed}).
  This implies that symplectic spectrum of the partial transposed Gaussian states of $\rho(V,d)^\sharp$ are same \cite{Simon1}.
  Note that, entanglement measure of formation is a function of the less symplectic spectrum of partial transposed Gaussian states if $a=b$ \cite{Weed,Ade3}, Gaussian states of $\rho(V,d)^\sharp$  have the same amount of entanglement under Gaussian measure of formation (The detailed definition of Gaussian entanglement of formation can be found in Appendix A for the convenience of readers). This shows elements of $\rho(V,d)^\sharp$ have the same amount of entanglement and coherence in the case $a=b$, respectively. That is, coherence and entanglement of $\rho(V,d)^\sharp$ are frozen simultaneously. It hints that there might have been a closed inner link between the measure of coherence and entanglement of formation.


 Theorem 3.2  and 3.3 are also key to characterize incoherent equivalence class of Gaussian states. It is clear that equivalence class of any pure coherent Gaussian state consists of some pure Gaussian states.
The most general pure Gaussian state $|\psi\rangle$ of one-mode is a displaced squeezed state obtained by the combined action of Weyl displacement operator
$${\widehat D}(\alpha)=e^{\alpha{\widehat a}^\dag-{\overline\alpha}{\widehat a}}, \quad \alpha\in{\mathbb C},$$ and the squeezing operator $${\widehat S}(\beta)=e^{\frac{1}{2}[\beta{{\widehat a}{^\dag}}^2-{\overline\beta}{\widehat a}^2]},\quad \beta\in{\mathbb C},$$
on the vacuum state $|0\rangle$ \cite{Olivx}: $$|\psi_{\alpha, \beta}\rangle={\widehat D}(\alpha){\widehat S}(\beta)|0\rangle.$$
The mean value and covariance matrix of $|\psi_{\alpha, \beta}\rangle$ are  $$ 2({\text Re}(\alpha), {\text Im}(\alpha)),$$

$$\left(\begin{array}{cc}
                                                                           {\text ch}(2|\beta|)+\cos\theta\ {\text sh}(2|\beta|)& \sin\theta\ {\text sh}(2|\beta|)\\
                                                                            \sin\theta\ {\text sh}(2|\beta|) & {\text ch}(2|\beta|)-\cos\theta\ {\text sh}(2|\beta)\end{array}\right),$$  where $\beta=|\beta|e^{i\theta}, {\text ch}(x)=\frac{e^x+e^{-x}}{2},\ {\text sh}(x)=\frac{e^x-e^{-x}}{2}$ are hyperbolic functions.
Denote $$\alpha=|\alpha|e^{i\gamma}, \alpha'=|\alpha'|e^{i\gamma'}, \beta'=|\beta'|e^{i\theta'},$$ by Theorem 3.3, a direct computation shows that $$\begin{array}{c}
            |\psi_{\alpha, \beta}\rangle\stackrel{\text IGO}{\sim}|\psi_{\alpha', \beta'}\rangle\\
            \Updownarrow\\
            |\alpha|=|\alpha'|, |\beta|=|\beta'|, \theta'-\theta=2(\gamma'-\gamma)-2k\pi\end{array}$$
 for some integer $k$.  It reveals explicitly geometric feature of incoherent equivalence class of displaced squeezed states \cite[Fig.1]{Adef}.

\vspace{0.1in}

{\it\bf IV Summary} \vspace{0.1in}

In this work, necessary and sufficient conditions for arbitrary multimode Gaussian states to be equivalent under incoherent Gaussian operations are obtained. It is shown that two coherent Gaussian states are incoherent  equivalence if and only if they are related by incoherent unitaries, if and only if  coherence of Gaussian states are frozen \cite{Ade4} under relative entropy measure.  Our results firstly provides an operational description of equivalent Gaussian states and so allow us to formulate a
simple criterion to decide whether Gaussian states are equivalent. Secondly, our results imply that all measures of coherence are frozen for an initial Gaussian state if and only if the relative entropy measure of coherence is frozen for the state. So this provides an entropy-based dynamical condition in which the coherence of an open quantum system is totally unaffected by noise.

Our results raise one interesting question naturally. How about the equivalence of coherent Gaussian states under stochastic incoherent Gaussian operations. The study may produce finitely many kinds of coherence and open a new door for deterministic Gaussian conversion protocols.

\vspace{0.1in}
{\it Acknowledgement---}
The authors thank the referees for their valuable suggestions which improved the presentation of the study.
We acknowledge that the research was  supported by NSF of China (12271452,11671332), NSF of
Fujian (2018J01006).

\vspace{0.1in}

{\it\bf Appendix A: Gaussian entanglement of formation}\vspace{0.1in}

For pure $N\times M$ Gaussian states $|\phi\rangle$, the Gaussian entanglement of formation is  defined as the von Neumann entropy of the reduced
states $\rho_{A,B}=Tr_{B,A}(|\phi\rangle\langle\phi|)$, i.e., $E_F(|\phi\rangle)=S(\rho_A)=S(\rho_B)$. The Gaussian entanglement of formation of mixed states  is defined as an infimum $$E_F(\rho)=\inf\{\ \sum_k p_kE_F(\phi_k)|\ \rho=\sum_k p_k|\phi_k\rangle\langle\phi_k|\}$$
over all possibly  convex decompositions (possibly continuous) of the state into pure states \cite{Weed}. In general, this optimization
is difficult to carry out. We only know the solution for two-mode symmetric Gaussian states  whose
covariance matrix is symmetric under the permutation of
the two modes, i.e., two diagonal elements are  equal when we write its covariance matrix in
the block form, where $E_F(\rho)$ is a function of the less symplectic spectrum of partial transposed Gaussian states.
\vspace{0.1in}

{\it\bf Appendix B: Proof of main results}\vspace{0.1in}

Proofs of all results in this paper are given in Appendix B.

{\bf Proof of Theorem 3.2.} By the definition of strictly incoherent Gaussian operations and monotonicity of coherence under all IGOs, it is clear that $(ii)\Rightarrow (iii)$. We will prove $(iii)\Rightarrow (i)$ and  $(i)\Rightarrow (ii)$ in the following.

``$(iii)\Rightarrow (i)$": For fixed orthonormal basis $(\{|n\rangle\}^{\infty} _{n=0})^{\otimes m} $ with positive integer $m > 1$, all Gaussian states are of the form
$\rho=\rho(V, d)=\rho ^{A_1A_2\ldots A_m}$, where $A_i$ denotes $i$th mode.  Furthermore, if $\rho=(\rho_{n_1n_2\cdots n_m,l_1l_2\cdots l_m})$ with $$\rho_{n_1n_2\cdots n_m,l_1l_2\cdots l_m}=\langle n_1 n_2\cdots  n_m|\rho |l_1l_2 \cdots l_m\rangle,$$
  \begin{equation}{\overline n_i}=\sum_{n_i}(\sum_{n_1n_2\cdots n_{i-1}n_{i+1}\cdots n_m} \rho_{n_1n_2\cdots n_m,n_1n_2\cdots n_m} )n_i,\end{equation} it is shown that $\mathcal C_R(\rho)=S(\rho||\delta)$ with some thermal state $\delta=\otimes_{i=1}^m\rho_{th}^{A_i}({\overline n_i})$ \cite{Xu1}.

Combining the definition of  relative entropy measure of coherence and monotonicity of relative entropy under completely positive and trace preserving mappings \cite{Lind},
 we have
$$\mathcal C_R(\Phi(\rho))\leq S(\Phi(\rho)||\Phi(\delta))\leq S(\rho||\delta)=\mathcal C_R(\rho).$$
From $\mathcal C_R(\Phi(\rho))=\mathcal C_R(\rho)$, it follows that
\begin{equation} \mathcal C_R(\Phi(\rho))=S(\Phi(\rho)||\Phi(\delta))= S(\rho||\delta)=\mathcal C_R(\rho).\label{2.1}\end{equation}
Recall that the strictly incoherent Gaussian operation $\Phi$  is described by a pair of operators
$(T, N)$, it performs on the Gaussian state $\rho(V, d)$ and
gets the Gaussian state with the mean value and the covariance matrix as:
 $$ d\mapsto Td, \hspace{0.1in} V\mapsto TVT^t+N.$$

Now, we demonstrate that there exists an IGO $\Psi$ such that $\Psi(\Phi(\delta))=\delta$. Noting that $\Phi(\delta)$ is an incoherent Gaussian state, we can assume  $\Phi(\delta)=\otimes_{i=1}^m\rho_{th}^{A_i}({\overline k_i})$. In order to use Petz recovery map of Gaussian systems \cite{Lami}, we have to check that $\Phi(\delta)$ is faithful, i.e., $V_{\Phi(\delta)}+i\Omega>0$, $V_{\Phi(\delta)}$ is the covariance matrix of $\Phi(\delta)$. By \cite{Xu1}, $$V_{\Phi(\delta)}=\bigoplus_{i=1}^m(2{\overline k_i}+1)I_2.$$
It is easy to check that $V_{\Phi(\delta)}+i\Omega>0\Leftrightarrow k_i> 0$ for $i=1,2,\cdots, m$.
Indeed, by (1) and (2), if $\overline k_i=0$ for some $i$, then
$$\begin{array}{ll}
\sum_{k_1k_2\cdots k_{i-1}k_{i+1}\cdots k_m} \Phi(\rho)_{k_1k_2\cdots k_m,k_1k_2\cdots k_m}=0,&\text{ if }k_i\neq 0,\\
\sum_{k_1k_2\cdots k_{i-1}k_{i+1}\cdots k_m} \Phi(\rho)_{k_1k_2\cdots k_m,k_1k_2\cdots k_m}=1, &\text{ if }k_i= 0.\end{array}$$
Thus $\Phi(\rho)$ has the form $|0^{A_i}\rangle \langle 0^{A_i}| \otimes \Phi(\rho)^{A_1A_2\cdots A_{i-1}A_{i+1}\cdots A_m}$. This deduces the covariance matrix $V_{\Phi(\rho)}$ of $\Phi(\rho)$ has the form $I_2\oplus V_2'$  which contradicts our assumption.
 Theorem 1 of \cite{Lami} shows that
the Petz recovery
map $\Psi$ is a  Gaussian operation with the following action:
$$\Psi:\left\{\begin{array}{ll}
V\mapsto T_{\Psi}VT_{\psi}^t+N_{\Psi}\\
d\mapsto T_{\Psi}d+d',\end{array}\right.$$
here $$T_{\Psi}=\sqrt{I+(V_{\delta}\Omega)^{-2}}V_{\delta}T^t(\sqrt{I+(\Omega V_{\Phi(\delta)})^{-2}}\  )^{-1}V_{\Phi(\delta)}^{-1},$$
$$N_{\Psi}=V_{\delta}-T_{\Psi}V_{\Phi(\delta)}T_{\Psi}^t,$$
$$d'=d_{\delta}-T_{\Psi}(Td_{\delta}+0).$$
From $V_{\delta}=\oplus_{i=1}^m (2\overline{n_i}+1)I_2$, $V_{\Phi(\delta)}=\oplus_{i=1}^m (2\overline{k_i}+1)I_2$ and $d_{\delta}=0$, it follows that $$T_{\Psi}=(\oplus_{i=1}^m\sqrt{(2\overline{n_i}+1)^2-1}I_2)T^t(\oplus_{i=1}^m \frac{ I_2}{\sqrt{(2\overline{k_i}+1)^2-1}}),$$ $$d'=0.$$
Note that $\Phi$ is a strictly incoherent Gaussian operation, we may write $$T=P_{\pi}\otimes I_2 \left(\begin{array}{ccc}
                           t_1O_1 & & 0\\
                               &\ddots &\\
                               0& & t_mO_m\end{array}\right),$$  $P_{\pi}$ is the permutation matrix corresponding to a permutation $\pi$ of $\{1,2,\cdots m\}$. One can check that
$$N_{\Psi}=V_{\Phi(\delta)}- T_{\Psi}V_{\delta}T_{\Psi}^t=\oplus w_i'I_2,$$ for some scalars $w_i'$.
It is evident that $T_{\Psi}\in{\mathcal T}_{2m}$. The remaining inequalities in the definition of IGOs are from complete positivity condition of $\Psi$. Thus $T_{\Psi}$, $N_{\Psi}$ satisfy the conditions of IGOs, and so $\Psi$ is a IGO with $\Psi(\Phi(\delta))=\delta$.

Next we claim that $\Psi(\Phi(\rho))=\rho$. A rotated Petz
map $\Psi^t$, for $t\in \mathbb R$,  is defined as
$\Psi^t(\omega)= \delta^{it}\Psi(\Phi(\delta)^{-it}\omega \Phi(\delta)^{it})\delta^{-it}$
with $\delta^{it} = \exp(it \log \delta)$ being understood as a unitary evolution according to the Hamiltonian
$\log \delta$ \cite{Wil15}. Obviously, $\Psi^t=\Psi$ if $t=0$. In \cite{MRDMA}, it is shown that
$$S(\rho||\delta)\geq S(\Phi(\rho)||\Phi(\delta))-\int_{\mathbb R}p(t) \log {\mathcal F}(\rho, (\Psi^{\frac t 2}(\Phi(\rho)))dt ,$$
where $p(t) = \frac{\pi}{2} (\cosh(\pi t) +1)^{-1}$ is a probability distribution parametrized by $t \in {\mathbb R}$, and  $\mathcal F$ denotes the quantum fidelity, defined as $\mathcal F(\omega, \tau) := \|\sqrt{\omega}\sqrt{\tau}\| _1 ^2$  for quantum
states $\omega$ and $\tau$. From equation (2), it follows that
\begin {align*}
 &\int_{\mathbb R} p(t) \log {\mathcal F}(\rho, (\Psi^{\frac t 2}(\Phi(\rho)))dt=0,\\
 &{\mathcal F}(\rho, (\Psi^{\frac t 2}(\Phi(\rho)))=1,\\
&\rho=\Psi^{\frac t 2}(\Phi(\rho)),  \forall t\in \mathbb R\\
&\rho=\Psi(\Phi(\rho)).\end{align*}

``$(i)\Rightarrow (ii)$": From (i),  there are matrices $T,S\in {\mathcal T}_{2m}$, and $N,N'$ such that
$$TVT^t+N=V',\ \ SV'S^t+N'=V.$$ Hence \begin{equation}STVT^tS^t+SNS^t+N'=V.\end{equation}
By the property of ${\mathcal T}_{2m}$, there are permutations $\pi,\pi'$ such that $P_{\pi}\otimes I_2T=\left(T_{ij}\right)$,
$P_{\pi'}\otimes I_2S=\left(S_{ij}\right)$,  where $\left(T_{ij}\right)$ and $\left(S_{ij}\right)$ are upper triangular blocks with the form
$$T_{ij}=\delta(i,f(j))t_iO_i,\ \ S_{ij}=\delta(i,g(j))s_iO_i'$$ here $f,g$ are functions from $\{1,\ldots,m\}$ to $\{1,\ldots,m\}$ with $f(j),g(j)\leq j,$ $$\delta(i,j)=\left\{\begin{array}{cc}
                            1,& i=j\\
                            0, & i\neq j\end{array}\right..$$ We need only show that $f(i)=g(i)=i$, $|t_i|=|s_i|=1$ and $\det O_i=\det O_i'=1$ for $i=1,\ldots,m$. Without loss of generality, we assume $T=\left(T_{ij}\right)$, $S=\left(S_{ij}\right)$.
Since there is at least one nonzero  off-diagonal element in every column of $V$ and $V'$, $T, S$ are invertible. Hence $f(i)=g(i)=i, i\in\{1,\ldots,m\}$ and so

$$T=\left(\begin{array}{ccc}
                           t_1O_1 & & 0\\
                               &\ddots &\\
                               0& & t_mO_m\end{array}\right), \quad
S=\left(\begin{array}{ccc}
                           s_1O_1' & & 0\\
                               &\ddots &\\
                               0& & s_mO_m'\end{array}\right),$$
 $$N=\left(\begin{array}{ccc}
                           \omega_1I_2 & & 0\\
                               &\ddots &\\
                               0& & \omega_mI_2\end{array}\right),  \quad
N'=\left(\begin{array}{ccc}
                           \omega_1'I_2 & & 0\\
                               &\ddots &\\
                               0& & \omega_m'I_2\end{array}\right).$$
We will show that $w_i=0$, $\det O_i=1$ and $|t_i|=1$.

Computing the diagonal blocks  of (3), we have, for $1\leq i\leq m$,
\begin{equation}
s_i^2t_{i}^2O'_{i}O_iV_{ii}O_{i}^t{O'}_{i}^t+w_is_i^2+w'_i
=V_{ii}. \end{equation} From the spectral theorem of positive operators, it follows that $s_i^2t_{i}^2\leq 1$.
Comparing blocks of $(i,j)$ $(1\leq i<j<m)$ positions in (3), we have
\begin{equation} s_i s_j t_{i} t_{j}O'_{i}O_iV_{ij}O_{j}^t{O'}_{j}^t
=V_{ij}.\end{equation} Note that, for each $i$, there is $j_i$ such that $V_{i,j_i}\neq 0$. Computing the trace norm of (5), we have $ |s_i s_{j_i }t_{i} t_{j_i}|=1$ and
$s_i^2t_{i}^2= 1$. From (4), we get $\omega_i=\omega'_i=0\  ( i=1,2,\ldots, m)$. Note that $\omega_i \geq |1- t_i^2\det O_i$, hence $\det O_i=1, |t_i|=1,\  (i=1,2,\ldots, m)$, as desired.

{\bf Proof of Theorem 3.3.} By the definition of IGOs, it is clear that $(ii)\Rightarrow (iii)$. The proof of $(iii)\Rightarrow (i)$ is almost verbatim from  $(iii)\Rightarrow (i)$ of Theorem 3.2. We need only show $(i)\Rightarrow (ii)$. We only treat $V\neq\lambda I_2$ and the other case can be treated similarily. Assume that there exists an IGO $\Phi$ such that $\Phi(\rho(V_1,d_1))=\sigma(V_2, d_2)$. By the definition of IGO, we can obtain
$$t^2 O VO^t+\omega I=V', \quad tOd_1=d_2, $$ where $O$ is a real orthogonal matrix, $\omega, t\in {\mathbb R},\  \omega\geq|1-t^2\det O|$. Similarily, there exist real orthogonal matrix $O'$ and real numbers $\omega', s$ with \begin{equation}s^2t^2O'OVO^tO'^t+\omega s^2 I_2+\omega' I_2=V,\end{equation}
$\omega'\geq|1-s^2\det O'|$. Suppose eigenvalues of $V$ are $\lambda_1$ and $\lambda_2$, it is clear that $\lambda_1\neq \lambda_2$.
By the spectral mapping theorem of positive operators, from (6), we have $$\{s^2t^2\lambda_1+\omega s^2+\omega', s^2t^2\lambda_2+\omega s^2+\omega'\}=\{\lambda_1,\ \lambda_2\}.$$
If $\left\{\begin{array}{ll}
           s^2t^2\lambda_1+\omega s^2+\omega' &=\lambda_2\\
           s^2t^2\lambda_2+\omega s^2+\omega'&=\lambda_1\end{array}\right.$, then $s^2t^2(\lambda_1-\lambda_2)=\lambda_2-\lambda_1$.
Thus $ts=0$ and so $V=\lambda I_2$ from (6). This is a contradiction. Therefore
$$\left\{\begin{array}{ll}
           s^2t^2\lambda_1+\omega s^2+\omega' &=\lambda_1\\
           s^2t^2\lambda_2+\omega s^2+\omega'&=\lambda_2\end{array}\right.$$ and  $|st|=1$. This implies $\omega=\omega'=0$.
Note that $\omega\geq|1-t^2\det O|$, hence $|t|=1$ and $\det O=1$ as desired.

\end{document}